\documentclass[letter]{aa} 
\usepackage{graphicx}
\usepackage{txfonts}
\usepackage{color}
\usepackage{natbib}
\usepackage{ulem}
\bibpunct{(}{)}{;}{a}{}{,} 
\newcommand{\BT}[1]{\textcolor{black}{#1}}
\newcommand{\SC}[1]{\textcolor{black}{#1}} 
\newcommand{\short}[1]{\textcolor{black}{#1}} 
\newcommand\kms{km\,s$^{-1}$}
\newcommand\DW{\textcolor{black}{DW}}

\begin{document}

   \title{ALMA discovery of a rotating SO/SO$_2$ {flow} in HH212}

  \subtitle{A possible MHD disk wind ?}

   \author{B. Tabone 
        \inst{1}, 
         S. Cabrit \inst{1,2}, E. Bianchi \inst{3,4}, 
        J. Ferreira \inst{2}, G. Pineau des Forêts \inst{1,5}, C. Codella \inst{3}, A. Gusdorf\inst{1},  F. Gueth\inst{6}, L. Podio\inst{3}, E. Chapillon\inst{6,7}}
\offprints{B. Tabone, \email{benoit.tabone@obspm.fr}}
\titlerunning{ALMA discovery of a rotating SO/ SO$_2$ wind in HH212}
\authorrunning{B. Tabone et al.}

   \institute{LERMA, Observatoire de Paris, PSL Research University, CNRS, Sorbonne Universit\'e, UPMC Univ. Paris 06, 75014 Paris, France 
                       \and
             Univ. Grenoble Alpes, CNRS, IPAG, 38000 Grenoble, France 
                       \and
             INAF, Osservatorio Astrofisico di Arceti, Largo E. Fermi 5, 50125 Firenze, Italy
             \and
             Università degli Studi di Firenze, Dipartimento di Fisica e Astronomia, Via G. Sansone 1, I-50019 Sesto Fiorentino, Italy
         \and
             Institut d’Astrophysique Spatiale, CNRS UMR 8617, Université Paris-Sud, 91405 Orsay, France 
         \and
             Institut de Radioastronomie Millimétrique, 38406 Saint-Martin d’Hères, France 
        \and
        OASU/LAB-UMR5804, CNRS, Université Bordeaux, 33615 Pessac, France }   
\abstract{
{}
{We wish to \short{constrain} the possible contribution of a magnetohydrodynamic disk wind (\DW) to the HH212 molecular jet.}
        {\short{We mapped the flow base with ALMA Cycle 4} 
        at 0\farcs13\,$\sim$\,60~au resolution
        and \BT{compared these observations} with 
        synthetic \DW\ predictions.}
{We identif\BT{ied}, in SO/SO$_2$\BT{,} a rotating flow \BT{that is} wider and slower than the axial SiO jet. The broad outflow cavity seen in C$^{34}$S is not carved by a fast wide-angle wind but by this slower agent.}
{Rotation signatures may be fitted by a \DW\ of a moderate lever arm launched out to $\sim$ 40 au with SiO tracing dust-free streamlines from 0.05-0.3 au. \SC{Such} a \DW\ could limit the core-to-star efficiency to $\leq$\,50\%.}}


   \maketitle
\vspace{10cm}

\section{Introduction}

 The question of angular momentum extraction from protoplanetary disks (hereafter PPDs) is fundamental in understanding the accretion process in young stars and the formation conditions of planets. Pioneering semi-analytical work, followed by a growing body of \BT{magnetohydrodynamic (MHD)} simulations, have shown that when a significant vertical magnetic field is present, {MHD} disk winds (hereafter \DW{s}) can develop that extract some or all of the angular momentum flux required for accretion \citep[see e.g.][and refs. therein]{2006A&A...453..785F,bethune2017,zhu2017}. The wind dynamics depend crucially on the disk magnetization, surface heating, and ionization structure, which are still poorly known in PPDs. Observing signatures of \DW{s} would thus provide unique clues to these properties. 
   
   Spatially resolved rotation signatures suggestive of a \DW\ were first reported in the intermediate velocity component ($V \simeq $ 50 \kms) surrounding the DG Tau optical atomic jet by \citet{2002ApJ...576..222B}. Their variation with radius was found \BT{to be} in excellent agreement with synthetic predictions for an extended \DW\ that extracts all of the accretion angular momentum out to a radius of 3~au \citep{2004A&A...416L...9P}. The inner regions of the same \DW\ could also explain the speed of the fast axial jet  \citep{2006A&A...453..785F}. However, rotation signatures in this faster component are less clear  \citep[eg.][]{louvet2016} \BT{because of} the limited spectral resolution and wavelength accuracy in the optical. Sub/mm interferometric observations do not have this limitation and have \BT{provided} clear evidence for flow rotation in several younger protostellar sources, although at lower speed\BT{s} than the axial jets, suggesting ejection from {$\sim 5-25$~au} in the disk \citep{launhardt2009, matthews2010, 2016Natur.540..406B, hirota2017}. Thermo-chemical models show that dusty DWs launched from this range of radii would indeed remain molecular despite magnetic acceleration \citep{2012A&A...538A...2P}. \BT{These models} would also reproduce all characteristics of the ubiquitous broad ($\pm 40$ \kms) H$_2$O line components revealed by {\it Herschel} in low-mass protostars \citep{yvart2016}. 
   
    More stringent tests of the \DW\ paradigm require high angular resolution. Using ALMA observations with an 8~au beam, \citet{2017NatAs...1E.152L} recently detected evidence for rotation in fast SiO jet knots from the HH212 protostar in the same sense as the rotating envelope. Assuming steady magneto-centrifugal launching and taking the observed gradient as a direct measure of specific angular momentum, \BT{these authors} inferred a launch radius of {$0.05^{+0.05}_{-0.02}$~au}, \BT{which suggested} that the SiO jet arises from the inner disk edge.
      Here we present Cycle 4 ALMA observations of the same source	at 0\farcs13$\sim$\,60~au resolution (for $d$ = 450 pc), \BT{which} reveal rotation in SO$_2$ and SO in the same sense as the SiO jet, but in a wider structure surrounding it. We compare the observations with synthetic predictions for extended \DW{s} to constrain the possible range of launch radii and magnetic lever arm, and \BT{we discuss the} major implications of our findings.

   \begin{figure*}[ht!]
   \centering
   \includegraphics[width=.95\textwidth]{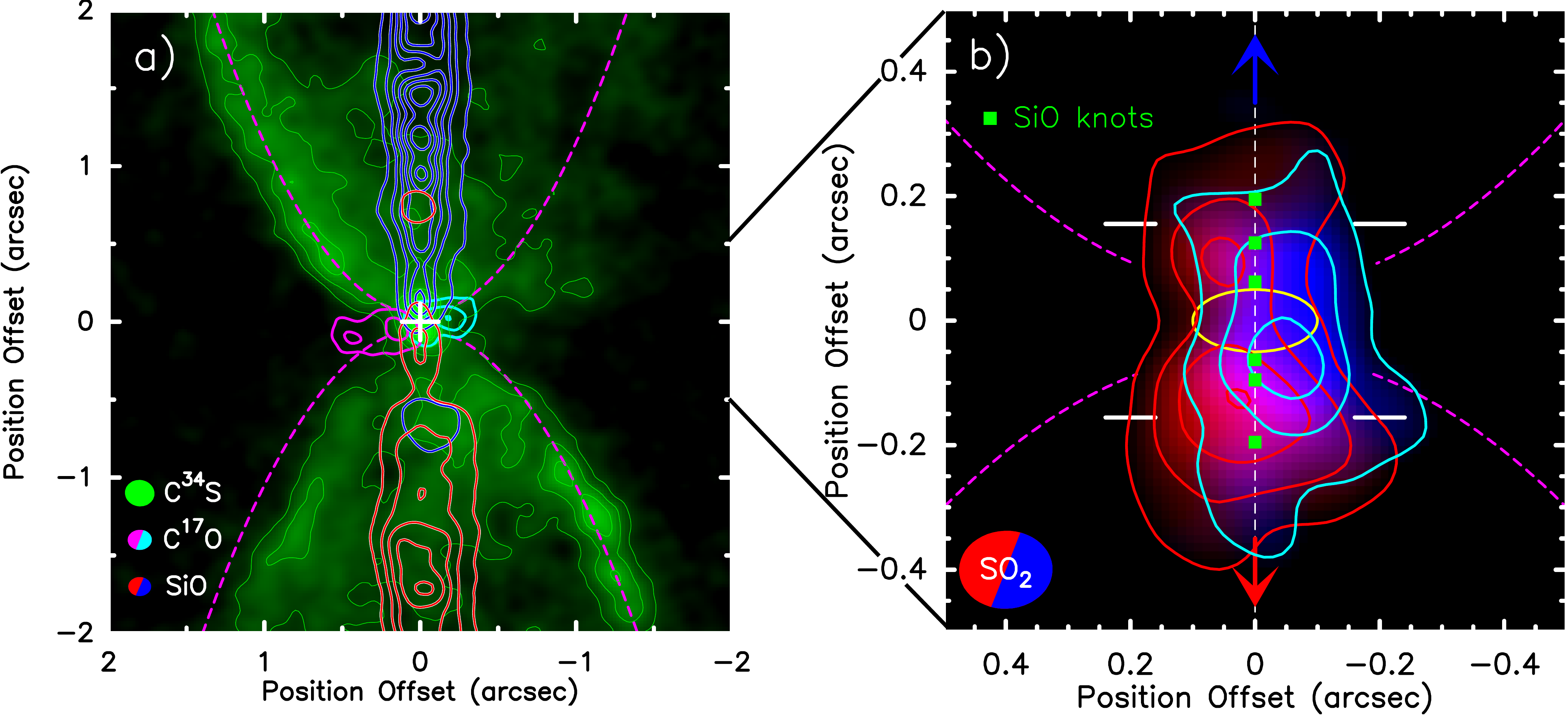} 
\caption{\textbf{a:} HH212 inner region viewed by ALMA Cycle 4. The SiO jet at $\mid V_{LSR} -V_{sys}\mid =$\,5-10\,\kms (blue and red contours), C$^{17}$O rotating \short{envelope} 
at $\mid$\,$V_{LSR} -V_{sys}$\,$\mid =$\,1.5 \kms (pink and turquoise)\BT{, and} cavity walls in C$^{34}$S at $\mid V_{LSR} -V_{sys}$\,$\mid \le$ 0.6\kms (green, with parabolic fits in dashed magenta) \BT{are shown}. 
First contour and step are $6 \sigma$ and $18 \sigma$ for SiO, $6 \sigma$ and $6 \sigma$ for C$^{17}$O, $4\sigma$ and $4\sigma$ for C$^{34}$S, where $\sigma$ is the rms noise level. 
The source is shown as a white cross and beam sizes are in the bottom left corner.
\textbf{b:} Zoom-in on blue and red SO$_2$ at $\pm2$ \kms\ \BT{shows} a slow outflow rotating in the same sense as the C$^{17}$O disk. {First contour and step are $7\sigma$.} White segments show the positions of PV cuts in Fig~\ref{fig:pv}. Green squares show inner SiO knots imaged by \citet{2017NatAs...1E.152L}; the yellow ellipse marks the centrifugal barrier radius at $r$$\sim$45~au and the height of the COM-rich disk atmosphere at $z \pm$20~au \citep{2017ApJ...843...27L}. Images are rotated such that the vertical axis corresponds to the jet axis at P.A = $22^{\circ}$.
}
   \label{fig:view}
   \end{figure*}  

\vspace{-0.3cm}
\section{Observations}

HH212 was observed in Band 7 with ALMA
between 6 \BT{October} and 26 \BT{November} 2016 (Cycle 4) using 44 
antennas of the 12-m array with a maximum baseline of 3 km.
{The SO$_2$($8_{2,6}-7_{1,7}$) line at 334.67335GHz was observed
with a spectral resolution of 1\kms and the
lines of SO($9_8-8_7$) 346.52848GHz, SiO(8-7) 347.33063GHz, C$^{34}$S(7-6) 337.39669GHz, and C$^{17}$O(3-2) 337.06113GHz with a spectral resolution of 0.1\kms\ (rebinned to 0.44 \kms).} Calibration was carried out following standard procedures under the CASA 
environment using quasars J0510+1800, J0552+0313, J0541--0211, and J0552--3627. Spectral line imaging was performed in CASA with natural weighting for C$^{34}$S {to increase sensitivity, resulting in a clean-beam  0.19''$\times$0.17'' (PA=-76$^{\circ}$), and with a R=0.5 robust factor for the other lines, resulting in a beam of 0.15'' $\times$ 0.13'' (PA $\sim$ -89$^{\circ}$). The rms noise level is $\sigma \sim$ 
1 mJy/beam in SO$_2$ in 1 \kms\ channels, and $\sigma \sim$1.5 mJy/beam in 0.44 \kms\ channels for the other lines.}
Further data analysis was performed using the
GILDAS\footnote{http://www.iram.fr/IRAMFR/GILDAS} package.
Positions are given with respect to the continuum peak at $\alpha({\rm J2000})$ = 05$^h$ 43$^m$ 51$\fs$41,
$\delta({\rm J2000})$ = --01$\degr$ 02$\arcmin$ 53$\farcs$17
(Lee et al. 2014) and velocities are with respect to a systemic velocity $V_{\rm sys}$ = 1.7 \kms  \citep{2014ApJ...786..114L}.

\section{Results and \BT{discussion}}
   
\subsection{Evidence for a rotating wide-angle flow in SO and SO$_2$}
   
Figure~\ref{fig:view}a presents a view of the various components of the HH~212 outflow system from our Cycle 4 data\BT{. The} chemical stratification first noted by \citet{2014A&A...568L...5C} in Cycle 0 is even more striking: SiO traces the narrow high-velocity jet, while C$^{34}$S outlines the dense walls of a broad outflow cavity, and C$^{17}$O traces the rotating equatorial envelope \SC{and} disk. We find that the cavity walls may be fitted by a parabolic shape $z = r^2/a + 0.05\arcsec$ with $a=0.9\arcsec$ in the \BT{n}orth and $a=1\arcsec$ in the \BT{s}outh \citep[slightly less open than sketched in][]{2017ApJ...843...27L}.

    \begin{figure*}[ht]
            \includegraphics[width=0.3\textwidth]{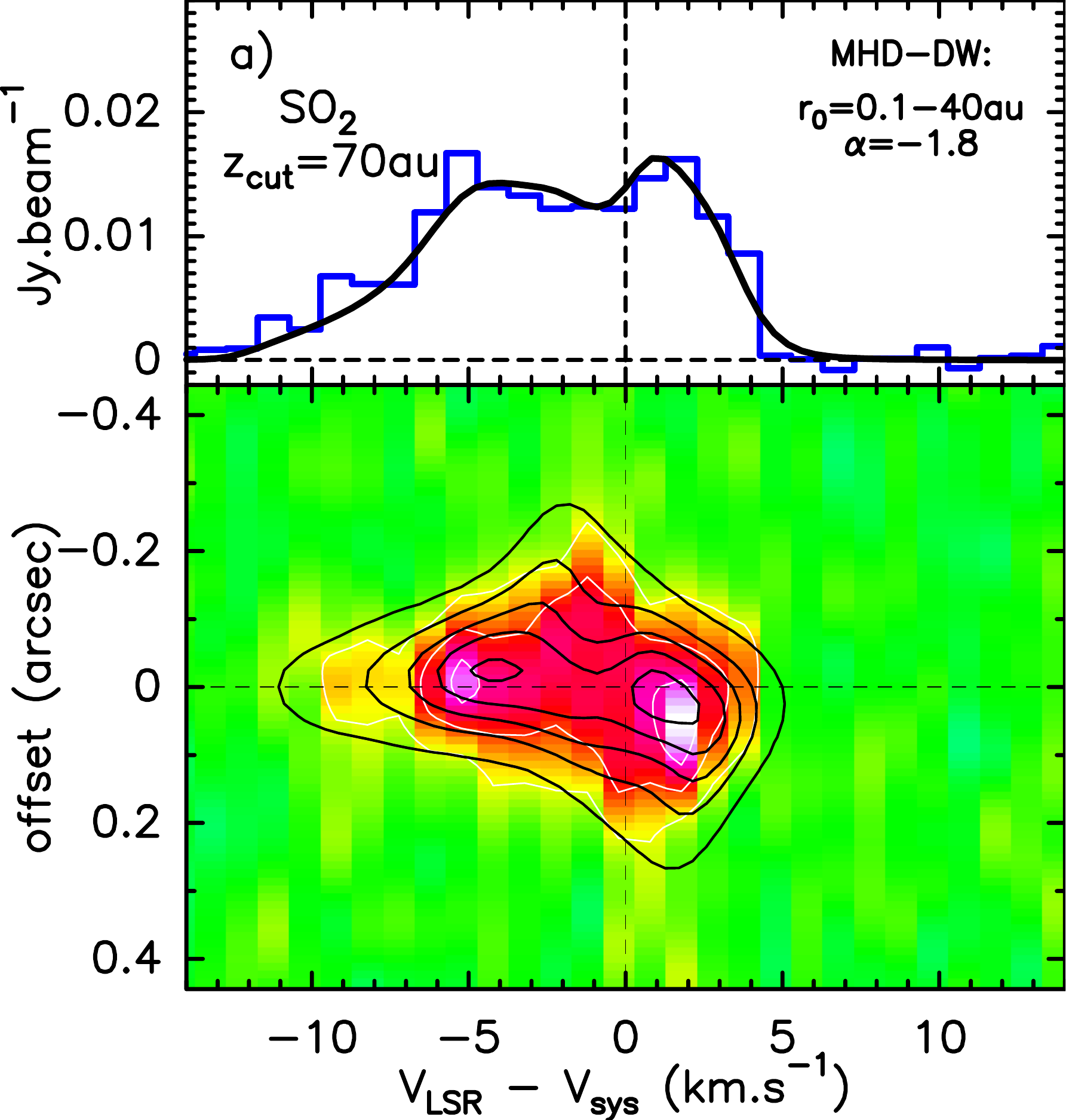}
         \includegraphics[width=0.3\textwidth]{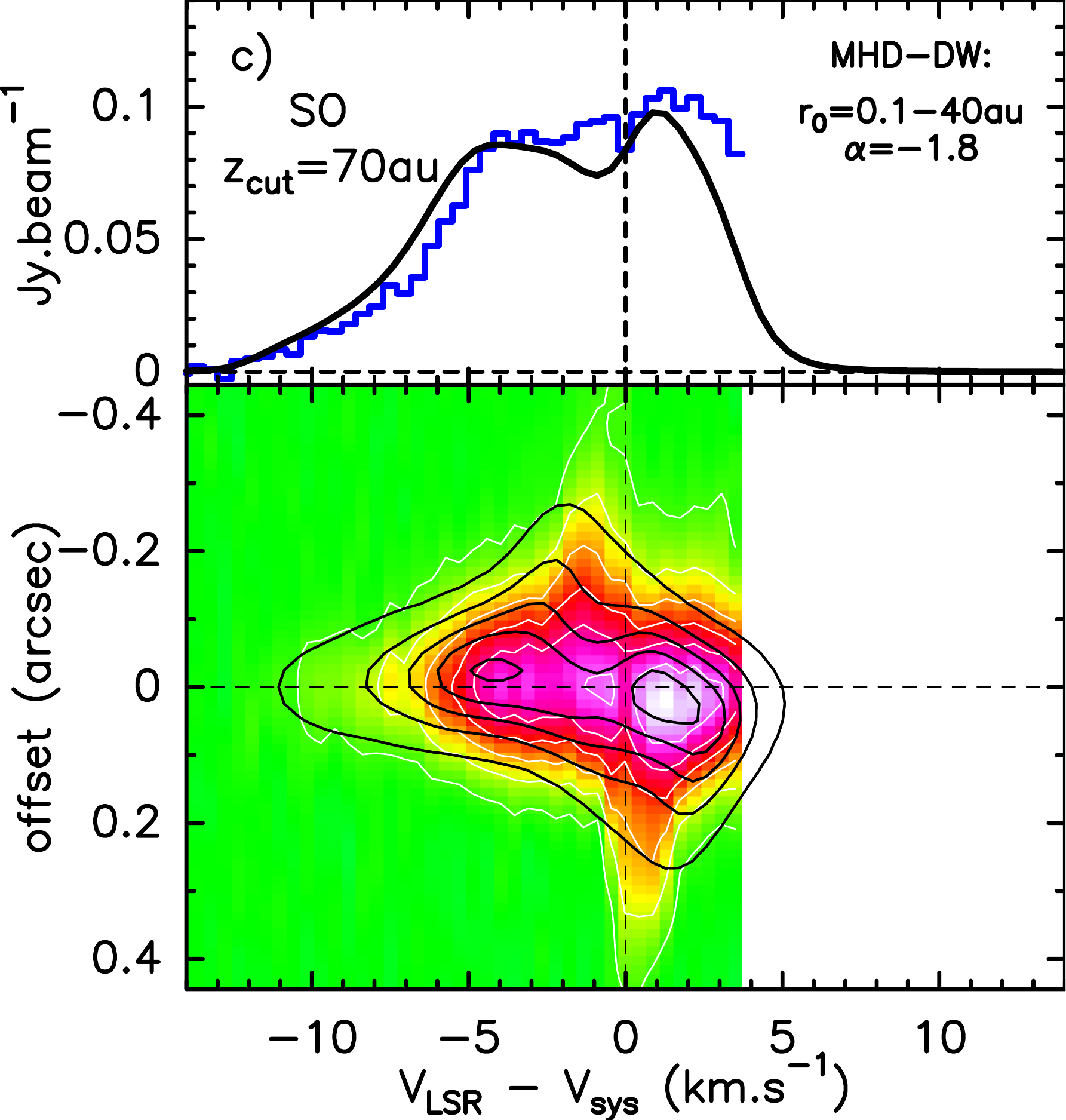}
     \includegraphics[width=0.3\textwidth]{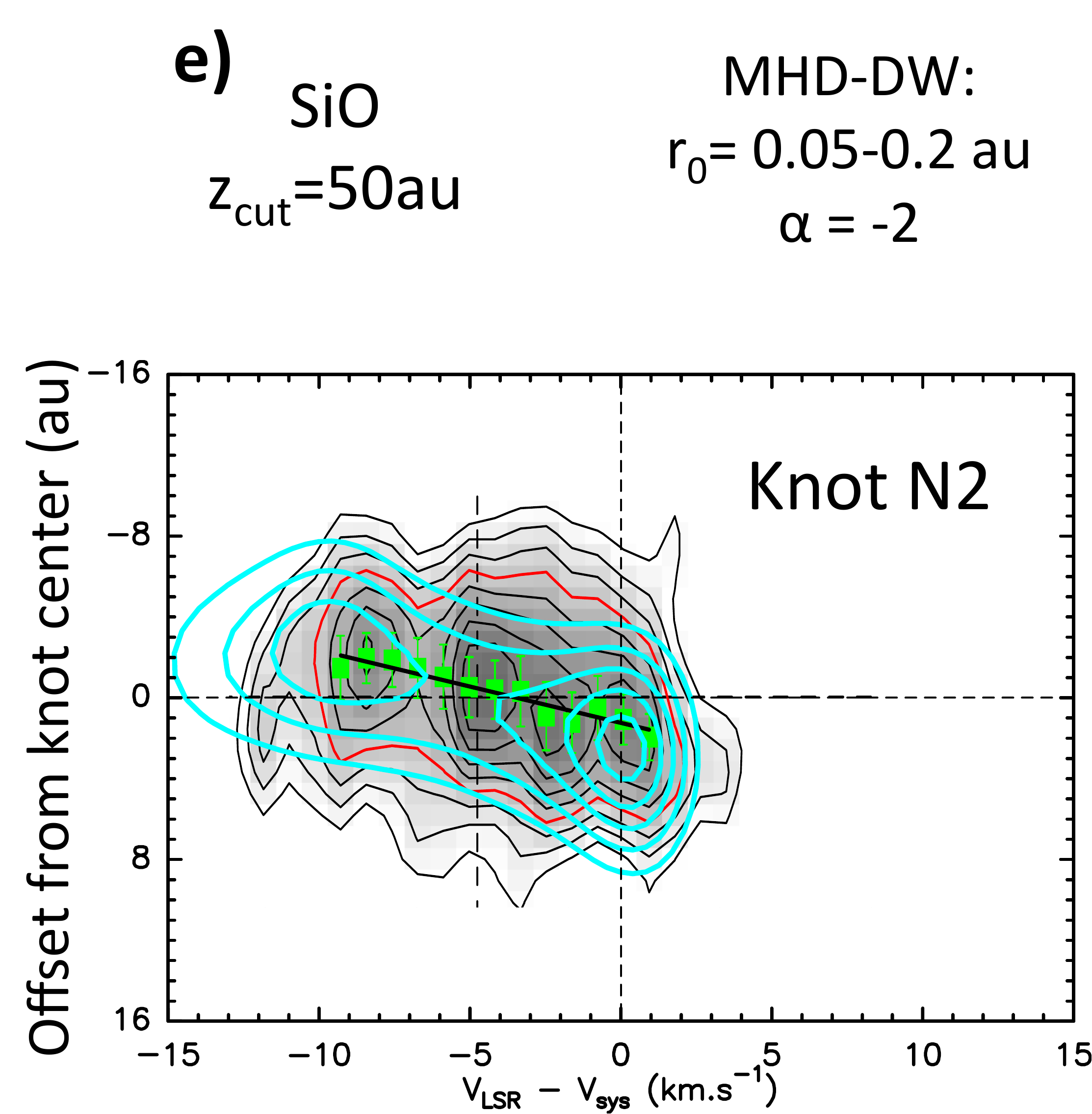} \\
        \includegraphics[width=0.3\textwidth]{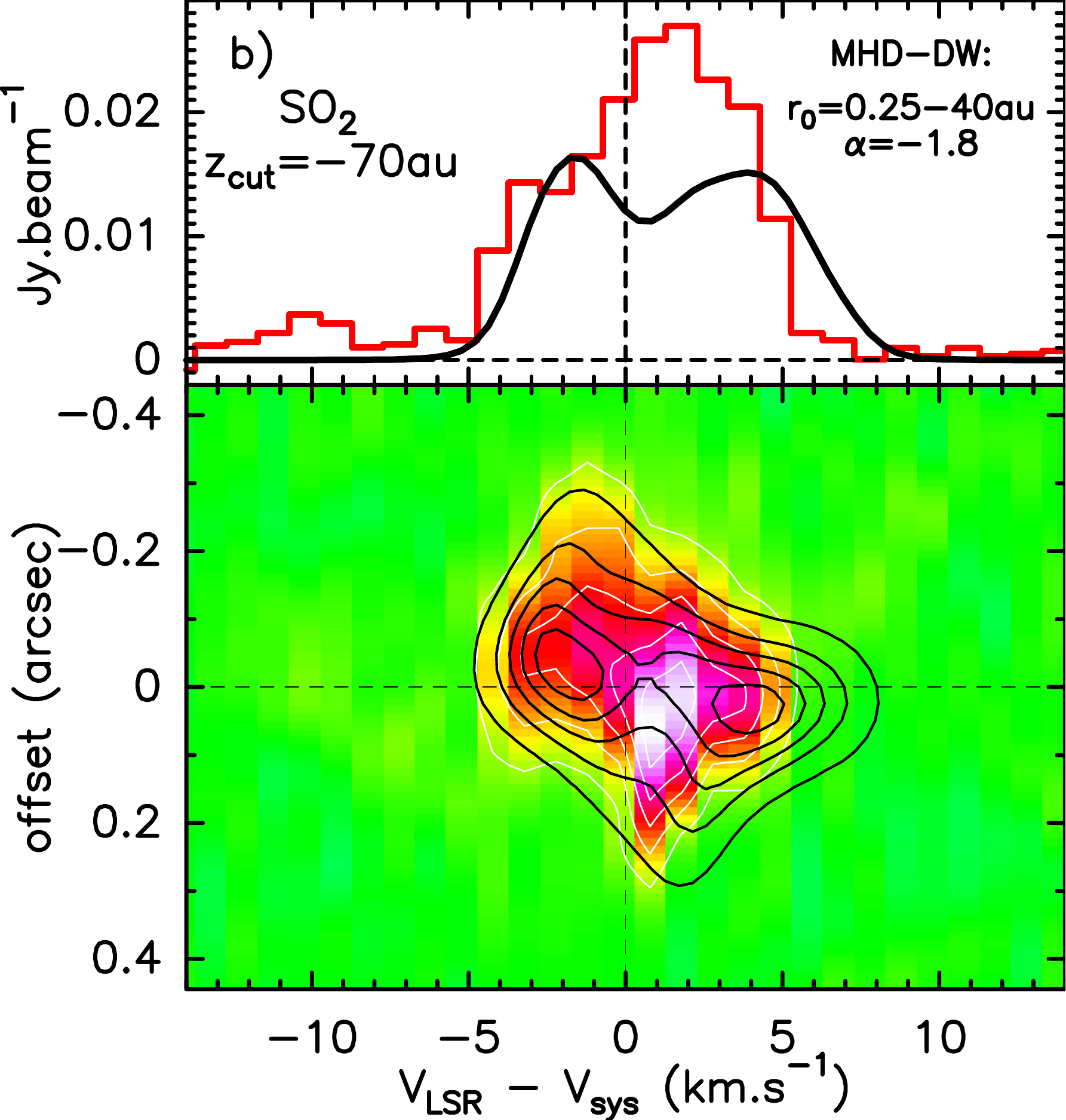} 
         \includegraphics[width=0.3\textwidth]{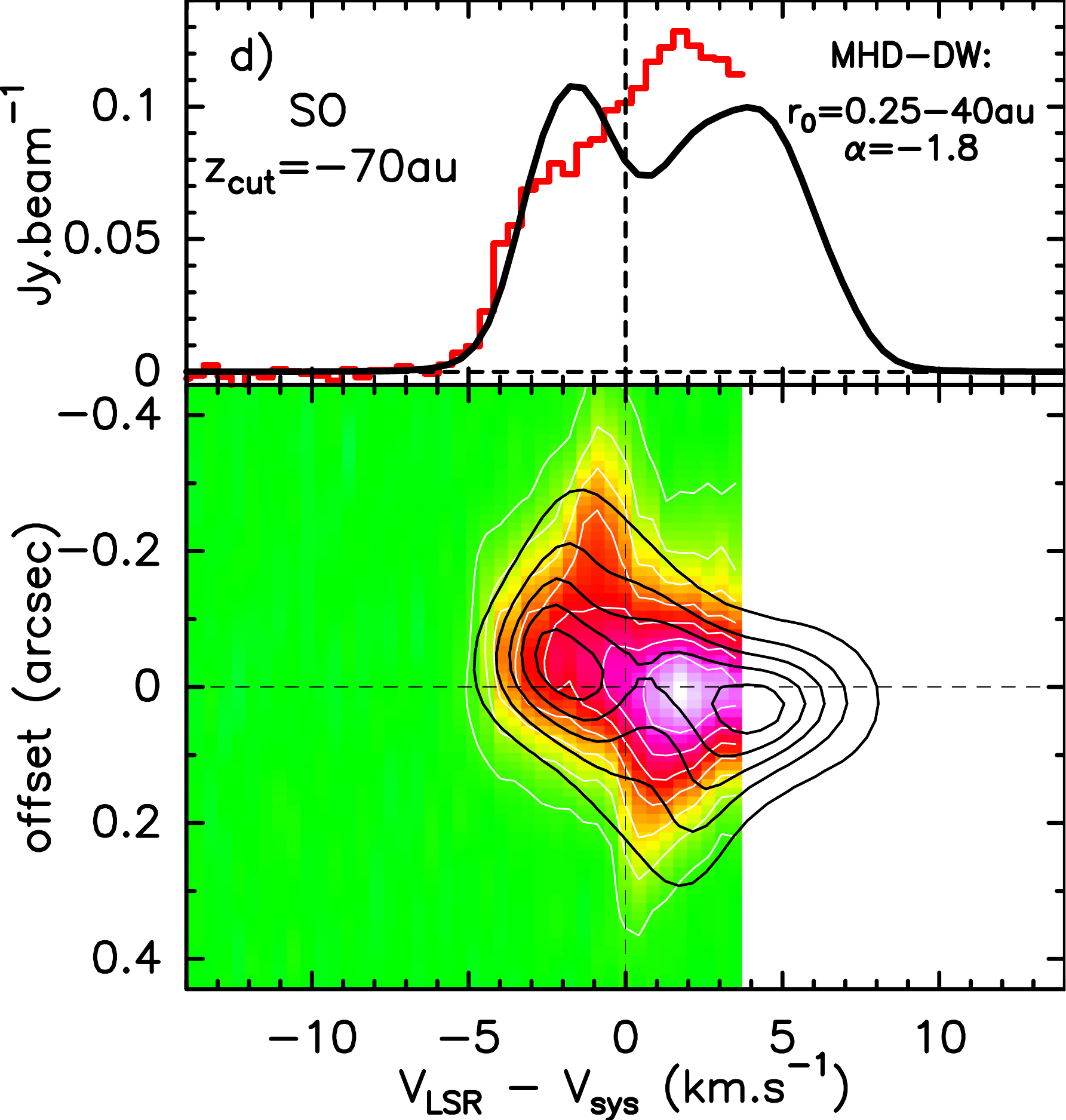}
        \includegraphics[width=0.3\textwidth]{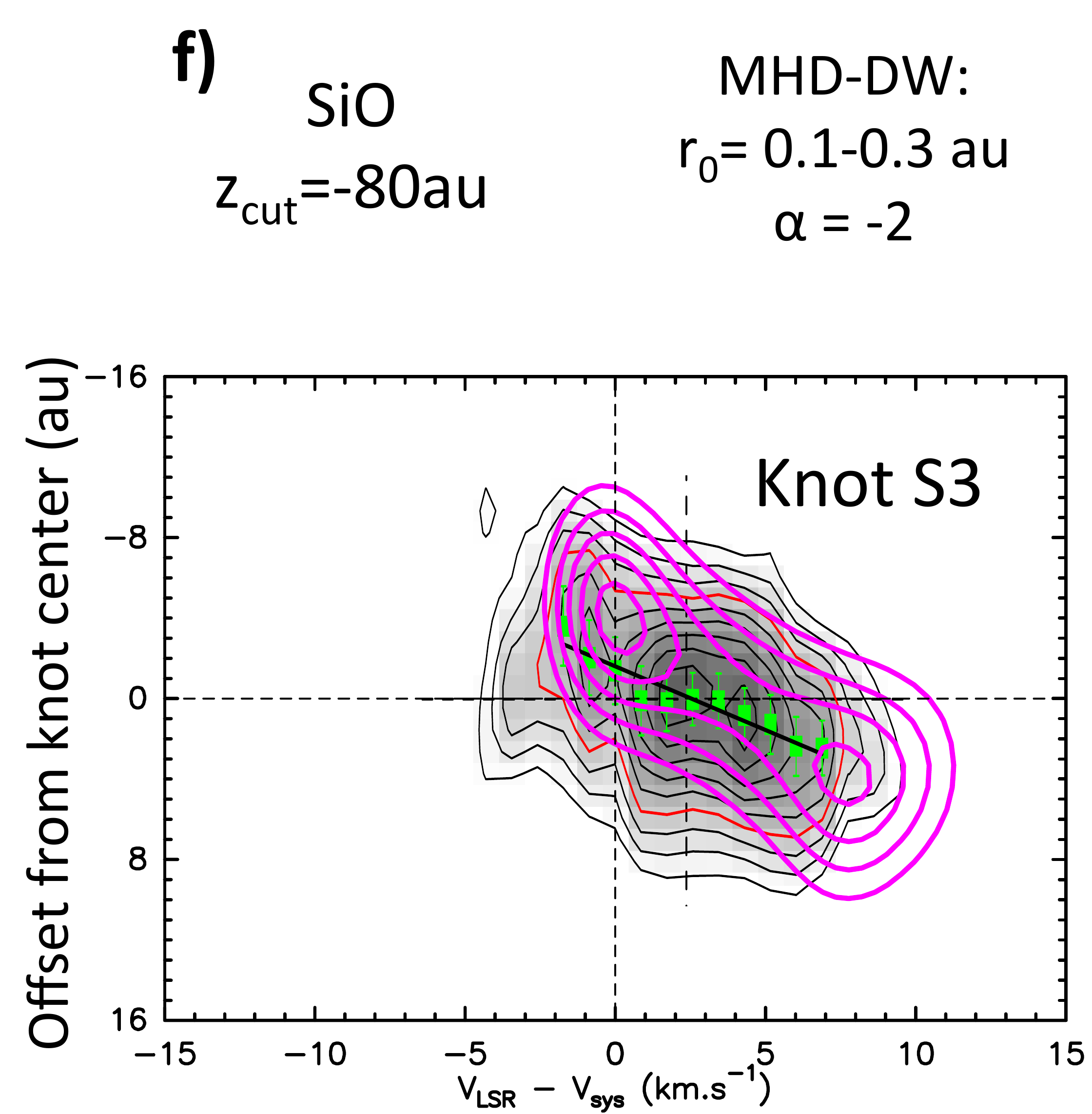}
      \caption{
      \textbf{a-d:} Observed on-axis line profiles (histograms) and transverse PV diagrams (color map and white contours)  of SO$_2$ (left) and SO (middle) at $\pm 70$~au across the \BT{northern} blue jet (top) and the southern red jet (bottom). The red wing of SO falls outside our spectral \BT{set-up}. A \DW\ model fit\BT{,} with $\lambda = 5.5$ and $\mathcal{W}=30$ (see text), is overplotted in black for parameters \BT{denoted} in each panel. 
      \textbf{e-f:} 
      PV diagrams \SC{observed by \citet{2017NatAs...1E.152L} across the SiO knots} N2 and S3  (\BT{grayscale} and black/red contours). Their measured centroids \SC{are shown} as green squares and fitted rotation gradient as a black line. The \DW\ model is \BT{overplotted} in cyan (top) or magenta (bottom) with parameters \BT{denoted} above each panel.}
     \label{fig:pv}
   \end{figure*}  

Figure~\ref{fig:view}b 
presents a zoom on SO$_2$ emission within 0.5\arcsec = 250 au of the source. This tracer, together with SO, was found to be abundant in the HH\,212 jet on \BT{a} larger scale \citep{2015A&A...581A..85P}. Our data resolve its bright emission near the base, revealing a rotation signature in the form of a transverse shift $\simeq \pm 0.05\arcsec$ between redshifted and blueshifted emission {at $\pm$ 2\kms\ from systemic}, in the same (\BT{east-west}) sense in both lobes and in the same sense as the envelope rotation in Fig.~\ref{fig:view}a.
The emission peaks at a typical distance of $0.1-0.15''$, well above the disk atmosphere 
at $z \simeq \pm 0.05$\arcsec\ traced by complex organic molecules  
\citep{2017ApJ...843...27L,bianchi2017} and extends out to 
$\pm$ 0.3\arcsec along the jet axis, indicating that this rotating material is outflowing. In the equatorial plane, the emission appears to originate from a region of typical radius $\simeq 0.1\arcsec$ = 45~au, \BT{i.e.,} similar to the \BT{centrifugal barrier} (herafter CB) estimated from HCO$^+$ infall kinematics, inside which the disk is expected to become \BT{K}eplerian  \citep{2017ApJ...843...27L}. 
Channel maps (see Fig.~\ref{chmaps-so2}) further show that this rotating outflow has an \BT{onion-like} velocity structure with
increasing width at progressively lower velocities, 
\BT{which} eventually \BT{fills up} the base of the cavity.
The SO emission has a similar behavior (see Fig.~\ref{chmaps-so}).
Figure~\ref{fig:pv} shows transverse position-velocity (PV) cuts and line profiles of SO$_2$ and SO at $\pm 70$~au from the midplane (beyond 1 beam diameter, to avoid any contamination by infall). Rotation is clearly apparent as a tilt in the PV cuts. \SC{The velocity decrease away from the jet axis is also visible.} The centroid velocity in \SC{on-axis} line profiles, $\sim$\,1-2\kms, yields a mean deprojected speed of $V_p \simeq 20-40$\kms\ for an inclination $i \simeq 87\degr$ \citep{1998ApJ...507L..79C}. {Hence, the outflow cavity is not carved by a fast wide-angle wind \SC{at $\sim 100$ \kms}, but by a slower component.}
\vspace{-0.2cm}
\subsection{Comparison with MHD disk wind models}

We compared the PV cuts and line profiles 
with synthetic predictions for steady-state, axisymmetric, self-similar \DW{s} from \BT{K}eplerian disks, calculated following the equations described in  \citet{2000A&A...353.1115C}. 
Two key properties of the MHD solution affect the predictions. \BT{First,} the magnetic lever arm parameter $\lambda \simeq (r_A/r_0)^2$, where $r_A$ is the Alfv{\'e}n radius along the streamline launched from $r_0$, which determines the extracted angular momentum and poloidal acceleration. \BT{Second,} the maximum widening $\mathcal{W} = r_{\rm max} / r_0$ reached by the streamline, which controls the flow transverse size. 
In order to limit the number of free parameters, we kept a fixed inclination $i = 87\degr$ and stellar mass $M_\star = 0.2M_\odot$ \citep{2017ApJ...843...27L}. The minimum and maximum launch radii, $r_{\rm in}$ and $r_{\rm out}$\BT{,} then determine the range of velocities in the wind (through the \BT{K}eplerian scaling). Since initial SO and SO$_2$ abundances at the disk surface are very uncertain, we did not compute the emissivity from a full thermo-chemical calculation along flow streamlines, as \BT{carried out} for H$_2$O by \citet{yvart2016}. \BT{Instead we} assumed a power-law variation with radius $\propto r^{\alpha}$ \BT{which allowed} us to investigate rapidly a broader range of parameters.  Synthetic data cubes were then computed assuming 
optically thin emission and a velocity dispersion of $0.6$~km s$^{-1}$ (the sound speed in molecular gas at 100~K), and convolved by a \BT{Gaussian} beam of the same FWHM as the ALMA clean beam. \BT{P}arameter $\alpha$ determines the relative weight of inner \BT{versus outer streamlines} and influences the predicted tilt in the PV. For a given MHD solution, \BT{the value of $\alpha$} is well constrained by the slope of the line profile wings.

We find that the \DW\ solution with $\lambda = 13$ used to fit the DG Tau jet in \citet{2004A&A...416L...9P} is too fast to reproduce the HH212 data\BT{; this solution} would require an angle from the sky plane of only {$0.5^{\circ}$}, outside the observed estimate of $4^{+3\degr}_{-1\degr}$ \citep{1998ApJ...507L..79C}. However, we could obtain a good fit for a slower MHD solution with $\lambda = 5.5$ and ${\mathcal{W}}$ = 30. 
While the emission peaks defining the tilt in PV diagram can be reproduced with $r_{\rm out} =$ 8~au, 
the more extended emission is better reproduced if we increase $r_{\rm out}$ to  the expected radius of the \BT{K}eplerian disk, \SC{namely 40~au}
\citep{2017ApJ...843...27L}. The corresponding best-fit predictions are superposed in black in Fig ~\ref{fig:pv}.
The value of $r_{\rm in}$ is constrained by the highest velocity present in the data; in the blue lobe, the extent of the blue wing suggests $r_{\rm in} \leq 0.1$~au. In the red lobe, our model fit is less good because the centroid is slower than in the blue by a factor 1.5-2. A slower solution with \BT{a} smaller lever arm (not yet available to us) would probably work better; numerical simulations show that it is indeed possible for a \DW\ to have asymmetric lobes \citep{fendt2013}. The value $r_{\rm in} \simeq 0.2$~au in Figs.~\ref{fig:pv}b,d is thus only \SC{illustrative}.
The model poloidal speeds at $z=70$\,au range from $\sim$\,100 to 2\kms\, for $r_o=$\,0.1 to 40\,au.

\vspace{-0.2cm}

Interestingly, we find that the same \DW\ solution that fits the SO and SO$_2$ PV cuts can also reproduce the rotation signatures 
\short{across} 
axial SiO knots at similar altitude, {if} SiO traces only inner streamlines launched from 0.05-0.1~au to 0.2-0.3~au. This is shown in Fig.~\ref{fig:pv}e-f, where our synthetic predictions, convolved by 8~au are compared with the ALMA SiO data of \citet{2017NatAs...1E.152L}. The predicted range of terminal speeds is 70-170\kms\BT{, which is} consistent with SiO proper motions.
Since the dust sublimation radius is also 0.2-0.3~au \citep{yvart2016}, SiO would be released by dust evaporation at the wind base.

  \subsection{Biases in {analytical} estimates of \DW\ outer launch radius}

An unexpected result is that our best fitting $r_{\rm out} \simeq 0.2-0.3$~au for SiO knots is 2-10 times larger than the $0.05^{+0.05}_{-0.02}$au estimated by \citet{2017NatAs...1E.152L} 
\short{with} the \citet{2003ApJ...590L.107A} formula, which is valid for all steady \DW{s}. Since the knots are resolved, this cannot be due to beam smearing as in the cases investigated by Pesenti et al. (2014). The same applies to our SO and SO$_2$ PV cuts, \BT{whereas} inserting the apparent velocity gradient in the Anderson formula would give $r_{\rm out} \sim$ 1~au instead of 40~au. 
We explored the reason \BT{for this discrepancy} and found that 
the superposition of {many} flow surfaces along the line of sight creates a shallower velocity gradient leading to strongly underestimate $r_{\rm out}$.  A detailed study of this effect\BT{,} which also leads to underestimating $\lambda$\BT{,} will be presented in Tabone et al. (in prep). 

\subsection{Further model tests and limitations}

\begin{figure}
   \centering
  \includegraphics[width=0.4\textwidth]{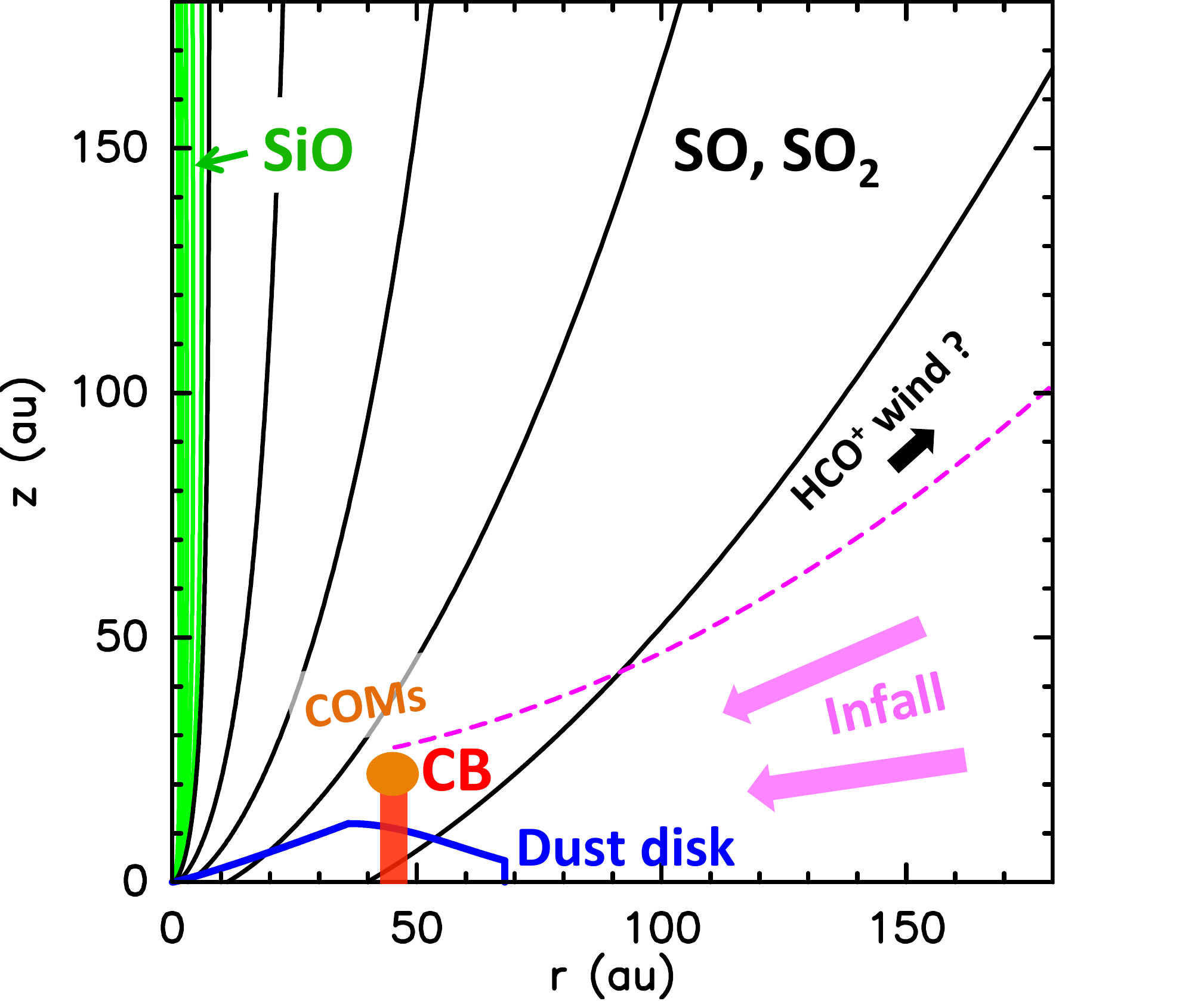} 
      \caption{Schematic view of the inner 180~au of the HH212 system following our {MHD disk-wind} \BT{modeling}. The streamlines rich in SiO are pictured in green, the wider component traced by SO and SO$_2$ is in black, for launch radii of 0.25, 0.9, 3, 11, \BT{and} 40~au. The magenta dashed curve shows the boundary of the cavity from Fig.~\ref{fig:view}. The dusty disk scale height from \citet{2017SciA....3E2935L} is \BT{indicated} in blue, the centrifugal barrier in red, and the COMs warm ring in orange \citep{2017ApJ...843...27L,bianchi2017}.}
         \label{fig:streamlines}
   \end{figure}

A strong test of the \DW\ picture would be to detect the predicted helical magnetic field structure, \BT{for example,} through dust polarization measurements with ALMA. Figure~\ref{fig:streamlines} plots the poloidal magnetic surfaces in our best-fit model. Beyond the Alfv\'en surface (at $z/r_0 \simeq 2$), a strong toroidal field develops. 
In the disk atmosphere at $z \sim$ 20~au, brightest in dust continuum, the ratio of {toroidal to poloidal magnetic field} $B_\phi/B_p$ ranges from 1.5 to 10, from outer to inner streamlines. Hence, polarization maps (if not dominated by dust scattering) should not show a pure "hourglass" geometry but be more toroidal closer to the axis. 

We also caution that our \DW\ modeling is only \BT{illustrative}, \BT{owing} to its simplifications. {Notably}, self-similarity cannot properly treat the effect of outer truncation. In reality, the shape and dynamics of the last \DW\ streamlines would be determined by pressure balance with the cavity and infalling envelope. As depicted in Figure~\ref{fig:streamlines}, they would thus open in the cavity more widely than predicted. 
This wider opening 
might explain the broader SO and SO$_2$ emission beyond the last model contour in PV diagrams and the slow HCO$^+$ wind noted by \citet{2017ApJ...843...27L}. The opposite pressure effect occurs near the equator, where the thick infalling stream would confine the streamlines inside the CB {more tightly than} \short{in our model.}
As shown in Fig.~\ref{fig:streamlines}, the heating resulting from this interaction might naturally explain the presence of a warm ring of COM emission 
close to the centrifugal barrier 
\citep{2017ApJ...843...27L,bianchi2017}.
\DW\ models {including these effects} remain to be developed.

\section{Conclusions}

Our Cycle 4 ALMA data reveal a rotating wide-angle flow in SO and SO$_2$ around the SiO jet with a mean speed of \short{$\sim 30$}\,\kms\ and an onion-like velocity structure filling in the base of the outflow cavity.
{Hence, the 
\short{cavity} is not carved by a fast wide-angle wind,
but by a slower component.}
\BT{This component} emerges from within and up to the centrifugal barrier, and contributes to remove excess 
angular momentum and mass from this region. 

The observed kinematics set tight constraints on stationary, self-similar MHD disk wind models. \BT{T}he lever arm parameter $\lambda$ should be \SC{$\la 5$,} smaller than in the atomic DG Tau rotating flow \citep[][]{2004A&A...416L...9P}, and the launch radii would range from 0.05 to $\sim$ 40~au, with SiO tracing only dust-free streamlines launched up to 0.2-0.3 au.
\short{If such a disk wind is} extracting most of the angular momentum required for disk accretion, it would \BT{be ejecting} 50\% of the incoming accretion flow\footnote{$\dot{M}_{\rm ej}/\dot{M}_{\rm acc}(r_{\rm out})$\,=\,$\left[1 - \left(r_{\rm in}/r_{\rm out}\right)^\xi \right]$ with $\xi \simeq 1/(2\lambda-2)$ when the wind braking torque dominates \citep{2000A&A...353.1115C}}. If \BT{this is} widespread among low-mass protostars, {as suggested by H$_2$O line profiles \citep{yvart2016}, this component} could strongly contribute to the low core-to-star efficiency $\simeq$ 30\% \citep[eg.][]{andre2010}. We nevertheless caution that our modeling is very idealized and probably not unique. Higher angular resolution and dust polarization maps with ALMA could provide powerful tests of \short{this} scenario.

Finally, an important side result of our study is that the 
apparent rotation gradient {strongly underestimates the actual outermost launch radius of an extended MHD {disk wind}.}
A detailed study of the {magnitude} of this effect, which applies beyond HH~212, will be presented in Tabone et al. (in prep).

\begin{acknowledgements}
We are very grateful to K.L.J. Rygl for her support with data reduction at the Bologna ARC node, and we thank the anonymous referee for useful comments. This paper makes use of the ALMA 2016.1.01475.S data (PI: C. Codella). ALMA is a partnership of ESO (representing its member states), NSF (USA)\BT{,} and NINS (Japan), together with NRC (Canada) and NSC and ASIAA (Taiwan), in cooperation with the Republic of Chile. The Joint ALMA Observatory is operated by ESO, AUI/NRAO\BT{,} and NAOJ. This work was supported by the Programme National “Physique et Chimie du Milieu Interstellaire” (PCMI) of CNRS/INSU with INC/INP and co-funded by CNES, and by the Conseil Scientifique of Observatoire de Paris. This research has made use of NASA's Astrophysics Data System.
\end{acknowledgements}
\vspace{-0.3truecm}
\bibliographystyle{aa} 
\bibliography{mybibli.bib} 

\begin{appendix}

\section{SO and SO$_2$ channel maps}
\label{app:channels}

   \begin{figure*}
   \centering
  \includegraphics[width=0.7\textwidth]{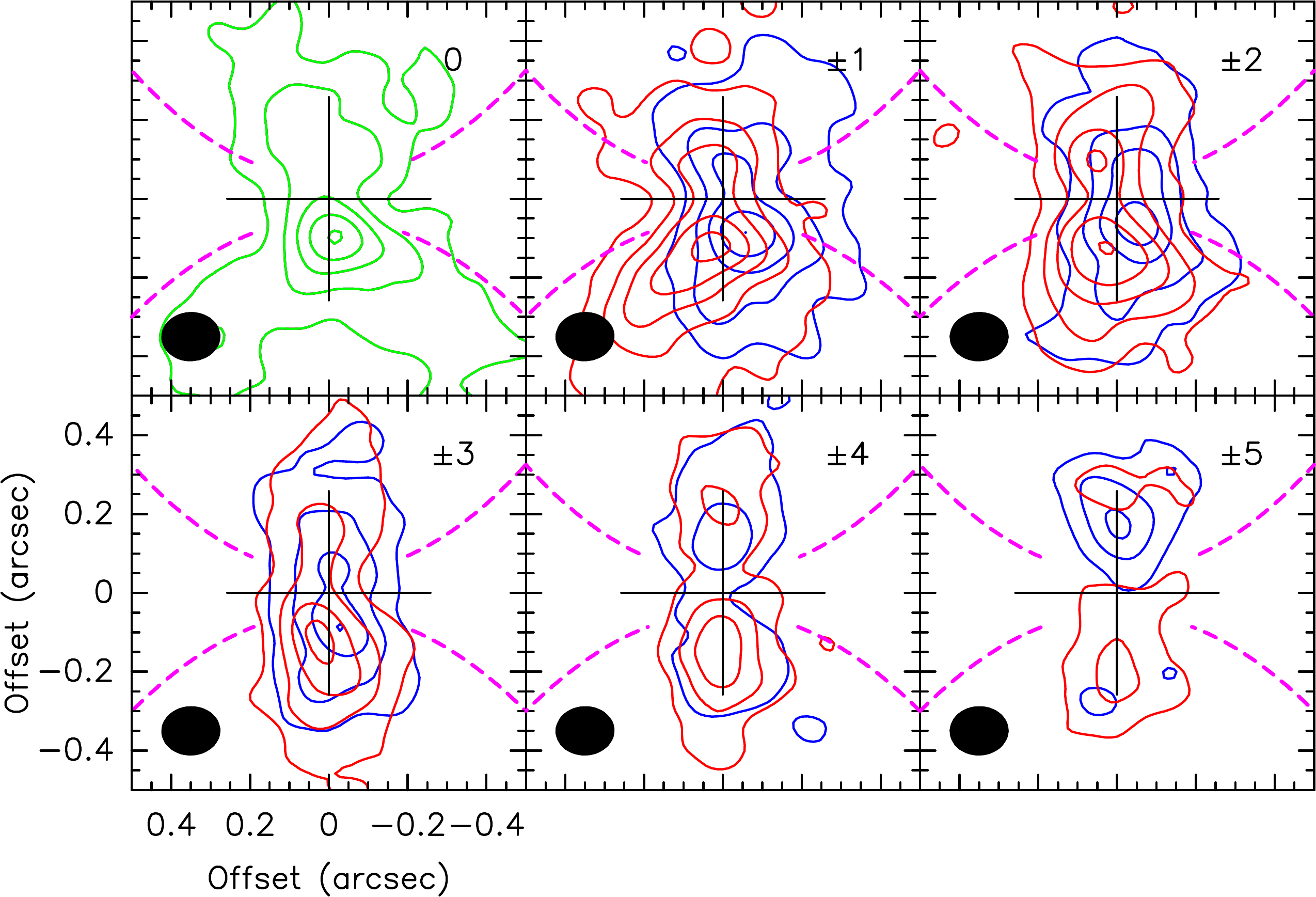} 
      \caption{Channel maps of continuum-subtracted SO$_2~8(2,6)-7(1,7)$  emission within $\pm $0.5\arcsec\ of the central source of HH212. The velocity offset from the systemic velocity ($V_{sys}=1.7$km s$^{-1}$) is indicated (in \kms) in the upper right corner with blue and red contours denoting blueshifted and redshifted emission. The channel width is 1 \kms. 
      First contour and steps corresponds to  4$\sigma$ and $6\sigma$ ($\sigma$=1mJy/beam)\BT{,} respectively. The C$^{34}$S cavity boundary from Fig.~\ref{fig:view} is drawn in magenta dashed lines.}
   \label{chmaps-so2}
   \end{figure*}  

   \begin{figure*}
   \centering
  \includegraphics[width=0.8\textwidth]{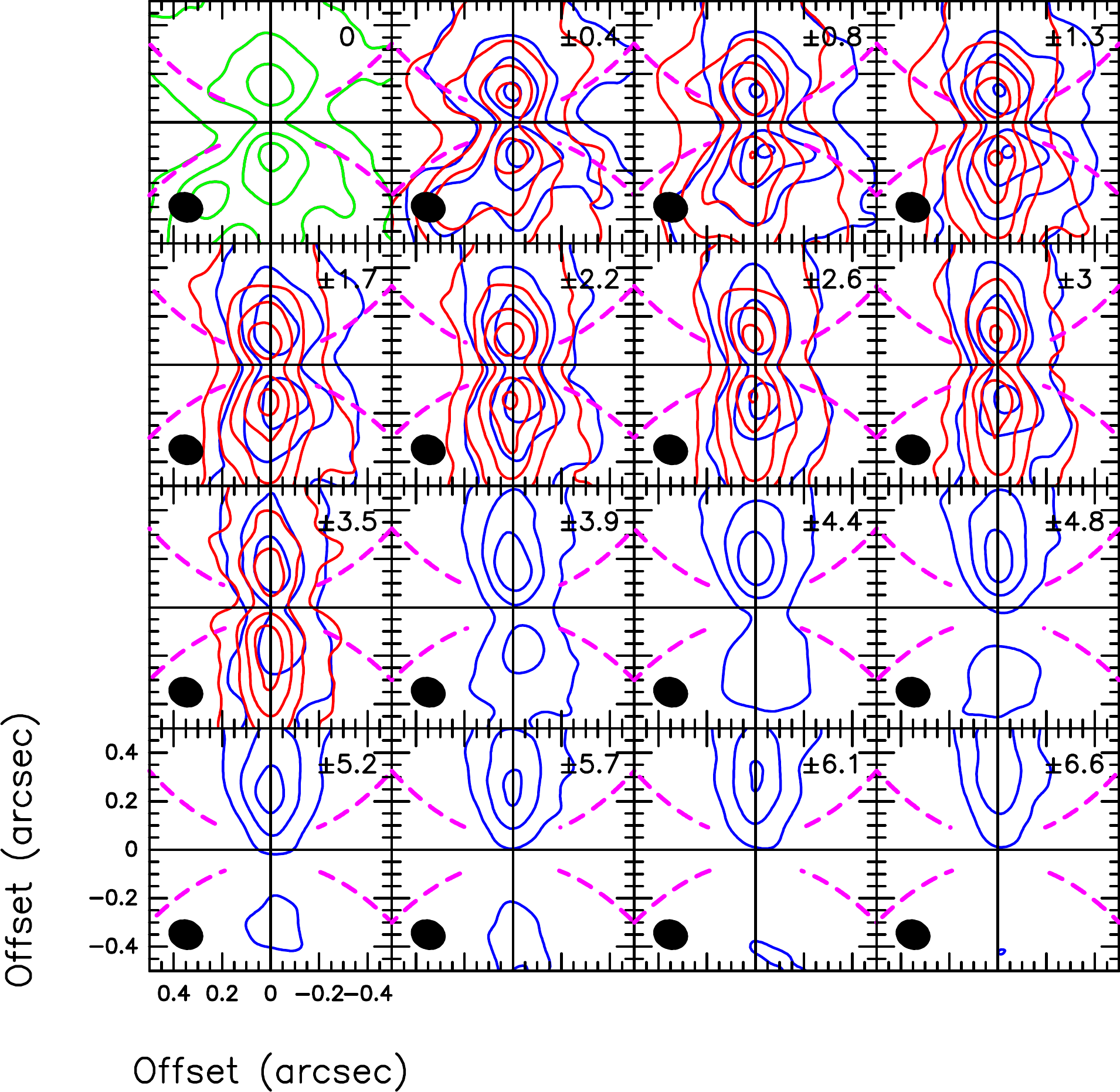} 
      \caption{Channel maps of {continuum-subtracted SO $9_8-8_7$ emission within $\pm $0.5\arcsec\ of the central source of HH212. The velocity offset from the systemic velocity ($V_{sys}=1.7$km s$^{-1}$) is indicated (in \kms) in the upper right corner with blue and red contours denoting blueshifted and redshifted emission. The channel width is 0.44 \kms. 
      First contour and steps corresponds to  6$\sigma$ and $16\sigma$ ($\sigma$ =1.7mJy/beam)\BT{,} respectively. The C$^{34}$S cavity boundary from Fig.~\ref{fig:view} is drawn in magenta} dashed lines.}
         \label{chmaps-so}
   \end{figure*}  

\end{appendix}

\end{document}